\documentclass[aps,prl,showpacs,twocolumn,amsmath,amssymb,10pt]{revtex4-1}
\usepackage{graphicx}
\usepackage{color}
\usepackage{siunitx}

\newcommand{\dd}{\,\mathrm{d}}
\newcommand{\mean}[1]{\left \langle #1 \right \rangle}

\begin{document}

\title{Effect of thermal noise on vesicles and capsules in shear flow}
\author{David Abreu and Udo Seifert}
\affiliation{{II.} Institut f\"ur Theoretische Physik, Universit\"at Stuttgart, 70550 Stuttgart, Germany}
\date{\today}

  \begin{abstract}
    We add thermal noise consistently to reduced models of undeformable vesicles and capsules in shear flow and derive analytically the corresponding stochastic equations of motion. We calculate the steady-state probability distribution function and construct the corresponding phase diagrams for the different dynamical regimes.  For fluid vesicles, we predict that at small shear rates thermal fluctuations induce a tumbling motion for any viscosity contrast. For elastic capsules, due to thermal mixing, an intermittent regime appears in regions where deterministic models predict only pure tank treading or tumbling.
  \end{abstract}

\maketitle

  {\sl Introduction.--}
    The dynamics of fluid vesicles and elastic capsules in linear shear flow has been extensively studied in the past decades (for reviews, see Refs. \cite{abka08,vlah09,bart11,fink11}). For fluid vesicles, the transition between the tank-treading regime (TT), in which the membrane rotates around the interior fluid at constant orientation, and the tumbling regime (TB), where the capsule rotates as a rigid body, is described with relatively good accuracy by the Keller-Skalak (KS) model \cite{kell82}, which assumes ellipsoidal vesicles of fixed shape. This assumption holds for moderate shear rates, while other dynamical regimes are observed if the vesicle deforms due to higher shear stresses \cite{kant06,made06,misb06,nogu07,lebe08,desc09a,kaou09a,faru10,zhao11,yazd12}. The Skotheim-Secomb (SS) model \cite{skot07} extends the KS model to capsules showing elastic behavior, such as the shape memory of red blood cells \cite{fisc04}. It captures many features observed in experiments \cite{abka07} and simulations \cite{kess08,bagc09,sui08,dods11}. For capsules, the TT motion exhibits a periodic oscillation of the orientation angle called swinging (SW). The TB regime also shows a periodic oscillation of the tank-treading angle. In addition, an intermittent regime (INT) appears for which both swinging and tumbling happen alternately. A thorough analysis of this intermittent regime is found in Ref. \cite{nogu09}, which suggests that intermittency should also exist for deformable capsules. However, analytical work \cite{fink11,vlah11} and simulations \cite{kess08,sui08,bagc09,walt11} tend to show that the intermittent regime may be only transient if capsule deformation is consistently taken into account.

    Since these objects are micrometer sized, it can be reasonably assumed that thermal fluctuations should have an influence on the dynamical transitions. Indeed, experiments on vesicles \cite{kant05,zabu11} show that they lead to sensible discrepancies from deterministic theoretical descriptions. However, most theoretical work that includes thermal noise has been done either for quasi-spherical vesicles in two dimensions (2D) \cite{fink08,me09} and three dimensions (3D) \cite{seif99} or through pure phenomenological models \cite{nogu05,nogu07,nogu10}. Thermodynamic consistency, however, imposes strong conditions on the stochastic equations of motion which we derive here in a minimal extension of the SS model. We then determine how such thermal fluctuations affect the phase diagram for vesicles and capsules. In particular, we show that under realistic conditions thermal effects should be observable.

\begin{figure}
	\centering
	\includegraphics[width=0.8\linewidth]{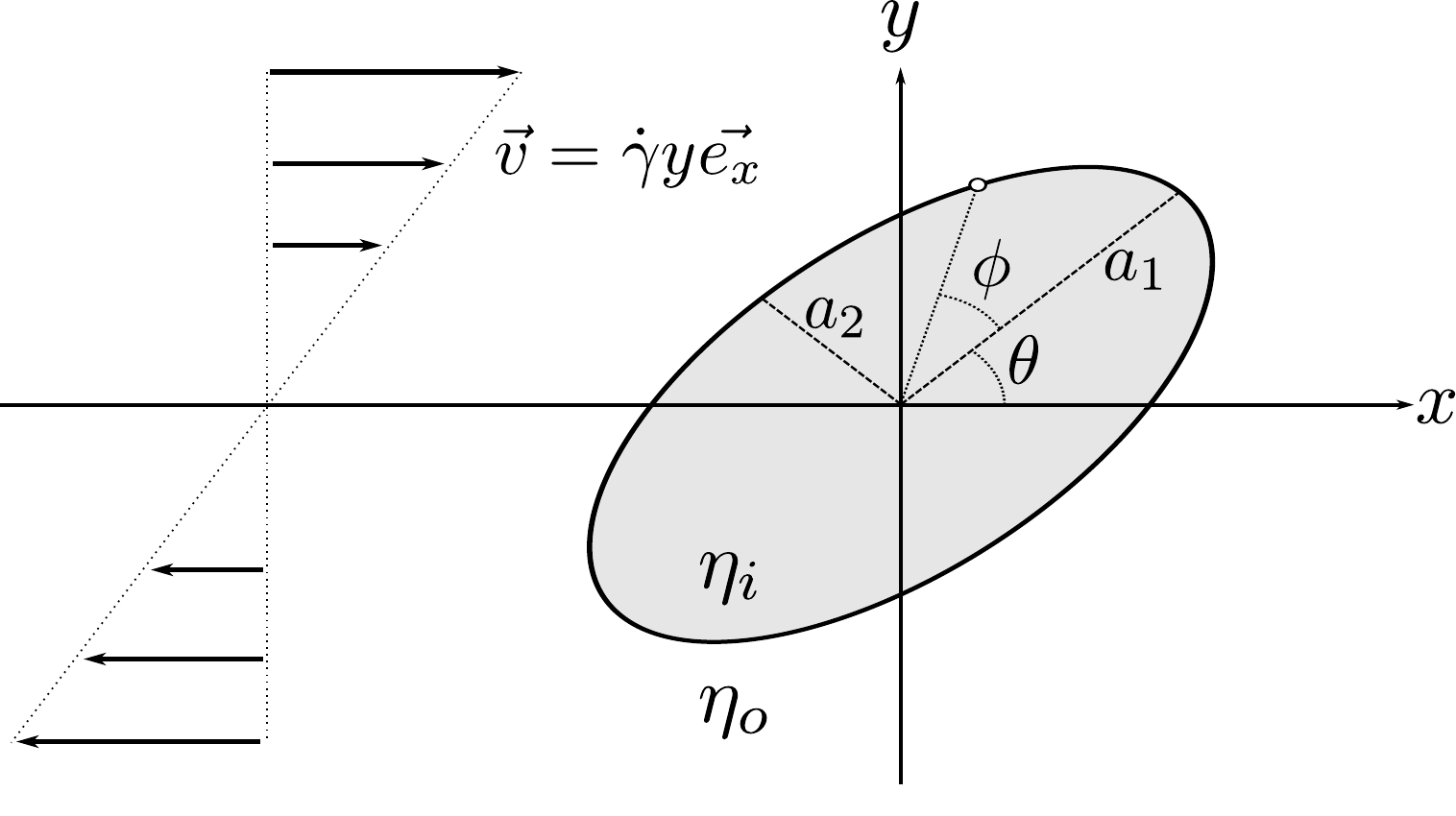}
	\caption{Schematic view of a capsule in shear flow.}
	\label{capsule}
      \end{figure}

  {\sl Reduced model.--}
    We describe the dynamics of an undeformable, ellipsoidal vesicle or capsule in linear shear flow as represented in Fig. \ref{capsule}. 
    The two degrees of freedom are the inclination angle $\theta$ of the particle relative to the middle plane of the shear flow and the phase angle $\phi$ describing the tank-treading motion of the outer membrane. Following Ref. \cite{skot07}, the balances of angular momentum and dissipated energy lead to the equations of motion 
      \begin{align}
	&\partial_t\theta = -\frac{\dot{\gamma}}{2}\left(1-\frac{z_1}{z_0}\cos(2\theta)\right) -\frac{\partial_t\phi}{z_0} + \zeta_\theta(t), \label{skottheta}\\
	&\partial_t\phi= -\frac{1}{z_1^2\lambda'}\left(\dot{\gamma}\frac{z_1}{z_2}\cos(2\theta) + \frac{E_0}{4\eta_oV}\sin(2\phi)\right) +\zeta_\phi(t) \label{skotphi},
      \end{align}
    which give the temporal evolution of $\theta$ and $\phi$ as functions of the shear rate $\dot{\gamma}$, the viscosity ratio $\lambda\equiv\eta_i/\eta_o$ between inner and outer viscosity, the main axes $a_1$, $a_2$, $a_3$ and the volume $V$ of the capsule, the elastic energy $E_0$ corresponding to the shape-memory effect, and geometrical quantities \footnote{$\lambda' \equiv \lambda+\frac{2}{z_2}-1,~ z_0\equiv\frac{1}{2}\left(\frac{a_1}{a_2}+\frac{a_2}{a_1}\right),~z_1\equiv\frac{1}{2}\left(\frac{a_1}{a_2}-\frac{a_2}{a_1}\right),~ z_2\equiv g'_3(\alpha_1^2+\alpha_2^2), ~ \alpha_i\equiv \frac{a_i}{(a_1a_2a_3)^{\frac{1}{3}}},~g'_3\equiv\int_0^\infty(\alpha_1^2+s)^{-\frac{3}{2}}(\alpha_2^2+s)^{-\frac{3}{2}}(\alpha_3^2+s)^{-\frac{1}{2}}\dd s$}. We quantify the influence of thermal fluctuations by adding the stochastic noises $\zeta_\theta(t)$ and $\zeta_\phi(t)$ similarly to Refs. \cite{nogu05,fink08}. They obey 
      \begin{equation}
	\mean{\zeta_i(t)}= 0, ~~~\mean{\zeta_i(t)\zeta_j(t')}=2D_i\delta_{ij}\delta(t-t'),
      \end{equation}
    where $\{i,j\}\in\{\theta,\phi\}$ and $D_\theta$ and $D_\phi$ are the respective diffusion constants. The rotational diffusion constant is           
      \begin{equation}
	D_\theta = \frac{k_\text{B}T}{V\eta_o} \frac{a_1^2g_1+a_2^2g_2}{4(a_1^2+a_2^2)},
	\label{Dtheta}
      \end{equation}
    which follows from the response of a rigid ellipsoid to a torque about the $z$ axis in a liquid of viscosity $\eta_o$ \cite{leal71}, $k_\text{B}$ being the Boltzmann constant, $T$ the temperature, and $g_1,g_2$ geometrical quantities \footnote{$g_1 \equiv\int_0^\infty(\alpha_1^2+s)^{-\frac{3}{2}}(\alpha_2^2+s)^{-\frac{1}{2}}(\alpha_3^2+s)^{-\frac{1}{2}}\dd s,~ g_2 \equiv\int_0^\infty(\alpha_1^2+s)^{-\frac{1}{2}}(\alpha_2^2+s)^{-\frac{3}{2}}(\alpha_3^2+s)^{-\frac{1}{2}}\dd s$}. To determine $D_\phi$, we use the fact that the shape-memory effect was introduced by adding the potential $E_0\sin^2\phi$ to the energy balance of the capsule. The fluctuation-dissipation relation in equilibrium ($\dot\gamma=0$) applied to Eq. \eqref{skotphi} then implies
      \begin{equation}
	D_\phi = \frac{k_\text{B}T}{V\eta_o}\frac{1}{4z_1^2\lambda'}.
	\label{Dphi}
      \end{equation}
    The diffusion coefficients \eqref{Dtheta} and \eqref{Dphi} depend on the inverse of the capsule volume $V$, i. e., the effects of thermal fluctuations become more important when the capsules are smaller. A small outer viscosity $\eta_o$ also amplifies thermal effects.

    Equations \eqref{skottheta} and \eqref{skotphi} form a set of Langevin equations which can be transformed into a single Fokker-Planck equation for the probability distribution $p(\theta,\phi,t)$ \cite{risk89}. This Fokker-Planck equation reads
      \begin{equation}
	\partial_t p(\theta,\phi,t) = -\partial_\theta j_\theta - \partial_\phi j_\phi,
	\label{fokkerplanck}
      \end{equation}
    with the currents
      \begin{align}
	j_\theta = -&\left[ \frac{\dot{\gamma}}{2}\right.\left(1 -B\cos(2\theta)\right) -\frac{E_0}{4\eta_oVz_0\lambda'z_1^2}\sin(2\phi) \nonumber \\
	&+\left. D_\text{eff} \partial_{\theta} - \frac{D_\phi}{z_0}\partial_\phi\right]p(\theta,\phi,t)
      \label{jtheta}
      \end{align}
    and
      \begin{align}
	j_\phi = -&\left[\frac{1}{z_1^2\lambda'}\right.\left(\dot{\gamma}\frac{z_1}{z_2}\cos(2\theta) + \frac{E_0}{4\eta_oV}\sin(2\phi)\right)  \nonumber \\
	& + \left.D_\phi\partial_\phi-\frac{D_\phi}{z_0}\partial_\theta\right]p(\theta,\phi,t),
	\label{jphi}
      \end{align}
    where we defined $B \equiv z_1/z_0+2/\lambda'z_0z_1z_2$ and 
      \begin{equation}
      	D_\text{eff} \equiv D_\theta + D_\phi/z_0^2.
      \end{equation}
    In the following, we will solve Eq. \eqref{fokkerplanck} in the steady state, i. e., for $\partial_t p(\theta,\phi,t) = 0$. 

    In the absence of shear flow ($\dot\gamma=0$), the equilibrium probability distribution becomes
      \begin{equation}
	p^\text{eq}(\theta,\phi) = \frac{1}{4\pi^2 I_0\left(\frac{E_0}{2k_\text{B}T}\right)} \exp\left(\frac{E_0}{2k_\text{B}T}(1-2\sin^2\phi)\right),
	\label{equildist}
      \end{equation}
    where $I_0(x)=1/\pi\int_0^\pi \exp(x\cos\theta)\dd\theta$ is a modified Bessel function of the first kind. This distribution does not depend on $\theta$ since no orientation is preferred. For fluid inelastic vesicles we have $E_0=0$, leading to a homogeneous distribution for both angles. For elastic capsules, the distribution shows two peaks at the angles $\phi=0$ and $\phi=\pi$ due to the shape-memory energy $E_0\sin^2\phi$. 

    In order to investigate the nonequilibrium steady states ($\dot\gamma>0$) and construct phase diagrams, we introduce the tumbling ratio \cite{kess09}
	\begin{equation}
	  \omega_\text{tu}\equiv\frac{\langle{\partial_t\theta\rangle}}{\langle{\partial_t\theta\rangle}+\langle{\partial_t\phi\rangle}},
	  \label{order}
	\end{equation}
    which plays the role of an order parameter: $\omega_\text{tu}=0$ corresponds to a pure tank-treading motion and $\omega_\text{tu}=1$ to pure tumbling, intermediate values corresponding to a dynamics in which both tumbling and tank treading take place. In the following, all numerical calculations are performed by setting $a_1=a_3=\SI{4}{\micro\meter}$ and $a_2=\SI{1.5}{\micro\meter}$, which corresponds approximately to the dimensions of a red blood cell \cite{abka07}. The outer viscosity is taken to be approximately that of water, $\eta_o=\SI{1}{\milli\pascal\second}$, and the temperature $T=\SI{293}{\kelvin}$.

  {\sl Fluid vesicles.--}
    Fluid vesicles do not exhibit shape memory, i. e., $E_0=0$, implying that the Fokker-Planck equation \eqref{fokkerplanck} does not contain any term depending explicitly on $\phi$. Therefore, we can write $p^\text{s}(\theta,\phi)=p^\text{s}(\theta)$. In addition, by averaging Eqs. \eqref{skottheta} and \eqref{skotphi}, one obtains
	\begin{equation}
	  \langle{\partial_t\phi\rangle} = -\frac{1}{\lambda'z_1z_2B}\left(2\langle{\partial_t\theta\rangle} +\dot{\gamma}\right),
	  \label{ttistb}
	\end{equation}
    such that the tank-treading frequency $\langle{\partial_t\phi\rangle}$ can be deduced from the tumbling frequency $\langle{\partial_t\theta\rangle}$. 
      
    Without thermal fluctuations, we recover the KS model for which two cases must be distinguished. If $B<1$, i. e., above the critical viscosity contrast $\lambda_c \equiv 1+2z_0/z_1z_2$, the solution of Eq. \eqref{fokkerplanck} is
      \begin{equation}
	p^\text{s}_\text{KS}(\theta,\lambda>\lambda_c)=\frac{\sqrt{1-B^2}}{2\pi(1-B\cos(2\theta))},
      \end{equation}
    which corresponds to a tumbling motion (TB) of the capsule around the $z$ axis. The mean tumbling frequency calculated from \eqref{skottheta} is equal to $\langle{\partial_t\theta\rangle} = -\dot{\gamma}\sqrt{1-B^2}/2$. Note that even though the capsule tumbles, the mean tank-treading velocity \eqref{ttistb} becomes zero only in the limit $\lambda\to \infty$, a fact already stressed in Ref. \cite{kell82}. The TB regime for vesicles therefore consists in a mixing between tumbling and tank treading. If $B\geq 1$, i. e., $\lambda \leq \lambda_c$, the inclination angle takes the stationary value
      \begin{equation}
	\theta_\text{TT} = \frac{1}{2} \arccos\left(\frac{1}{B}\right),
	\label{ttkelskal}
      \end{equation}
    while the membrane exhibits a tank-treading motion (TT) around the interior of the capsule. Here we have $\langle{\partial_t\theta\rangle} = 0$ and $\langle{\partial_t\phi\rangle} = -\dot{\gamma}/\lambda'z_1z_2B$, which corresponds indeed to pure tank treading without tumbling. This transition between tumbling and tank treading happens at the critical viscosity ratio $\lambda_c$ which depends only on the geometry of the vesicle but not on the shear rate $\dot{\gamma}$. The order parameter \eqref{order} is $\omega_\text{tu}=0$ for TT, and $0<\omega_\text{tu}\leq1$ for TB.

\begin{figure}
	\centering
	\includegraphics[width=0.99\linewidth]{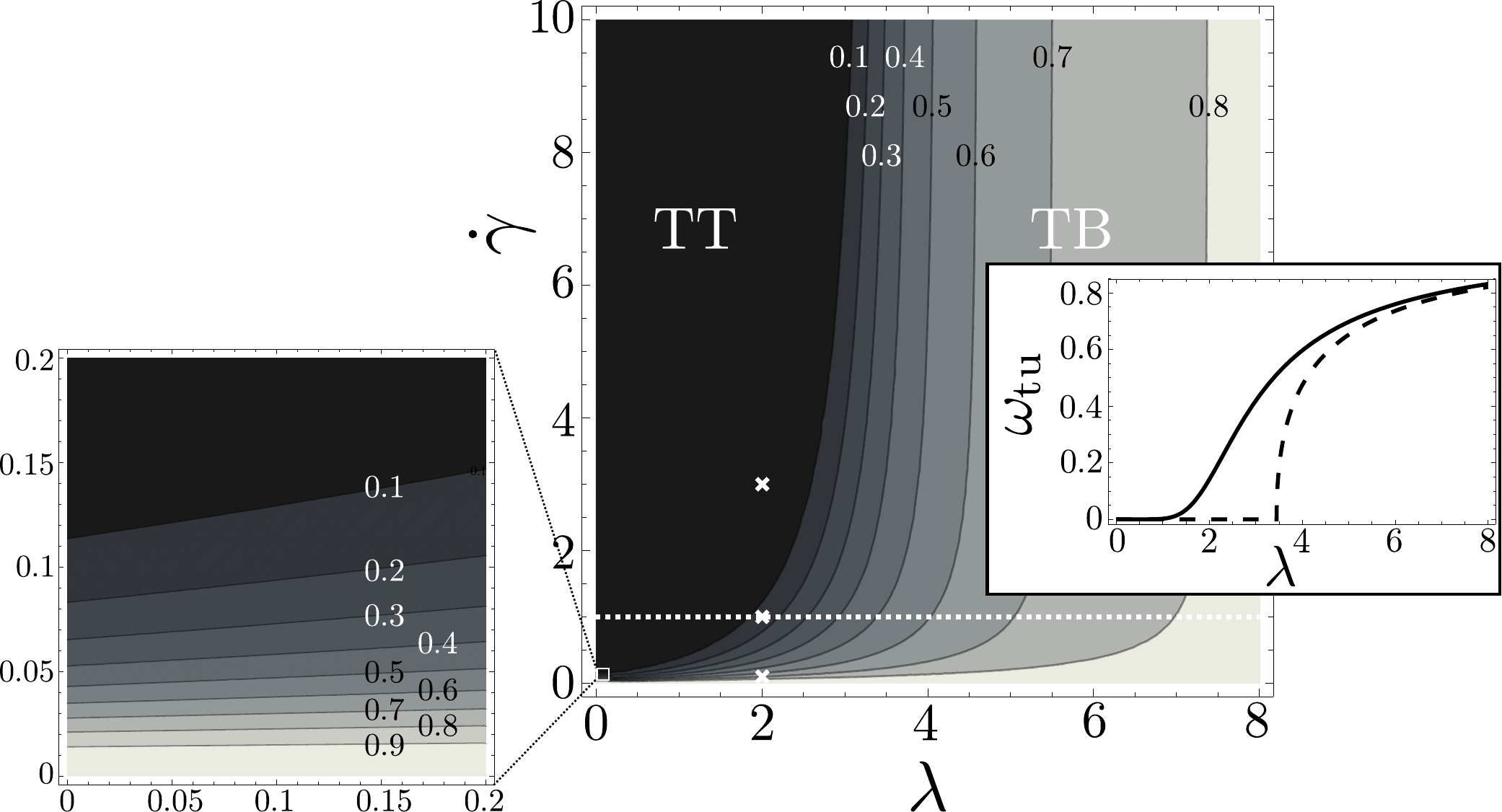}
	\caption{Phase diagram for the KS model with thermal noise as a function of $\dot\gamma$ and $\lambda$. Inset: $\omega_\text{tu}$ along the white dotted line ($\dot\gamma=\SI{1}{\second^{-1}}$), and the dashed line corresponds to the deterministic KS model. An enlargement of the bottom left corner is also shown.}
	\label{ves_pd_velo}
      \end{figure}

    If we include thermal noise, the Fokker-Planck equation \eqref{fokkerplanck} can still be easily solved as
      \begin{equation}
	p^\text{s}(\theta) = \frac{1}{\cal N}\int_0^\pi\dd\theta' \exp\left( \frac{\dot{\gamma}}{2D_\text{eff}} \left[ \theta'-B\sin\theta'\cos(\theta'+2\theta) \right] \right),
	\label{psnoem}
      \end{equation}
    where 
      \begin{equation}
	{\cal N} = 2\pi\int_0^\pi\exp\left(\frac{\dot{\gamma}}{2D_\text{eff}}\theta'\right)I_0\left(\frac{\dot{\gamma}}{2D_\text{eff}}B\sin\theta'\right)\dd\theta'
      \end{equation}
    is the normalization constant and $I_0$ the Bessel function defined after Eq. \eqref{equildist}. In this case, the stationary probability distribution \eqref{psnoem} depends on the shear rate $\dot{\gamma}$, which was not the case without noise. The mean tumbling frequency
      \begin{equation}
	\langle{\partial_t\theta\rangle} = -2\pi\frac{D_\text{eff}}{\cal N}\left( \exp \left[ \frac{\dot{\gamma}\pi}{2D_\text{eff}}\right] -1 \right)
	\label{tumbfr}
      \end{equation}
    is never exactly zero, such that $\omega_\text{tu}\neq0$ always. However, as can be seen in the inset of Fig. \ref{ves_pd_velo} showing the tumbling ratio as a function of $\lambda$ for $\dot\gamma=\SI{1}{\second^{-1}}$, it decays exponentially in the black region. We can thus still effectively identify a TT regime, defined here by $\omega_\text{tu}<0.1$, in which tumbling becomes rare. The contour plot of Fig. \ref{ves_pd_velo}, which represents $\omega_\text{tu}$ as a function of $\lambda$ and $\dot\gamma$, can therefore be interpreted as a phase diagram, where TT occurs in the black region and TB elsewhere.

    The first effect of thermal noise is to smooth out the TT-TB transition, which happens for smaller viscosity contrasts than in the deterministic case, as can be seen in the inset of Fig. \ref{ves_pd_velo}. Moreover, in this case, this transition depends not only on $\lambda$ but also on $\dot\gamma$. In the limit of high shear rates, fluctuations are suppressed  and we recover the critical viscosity contrast of the KS model. On the other hand, the critical viscosity contrast goes to $0$ for low but non-zero shear rates. In fact, as the enlargement in Fig. \ref{ves_pd_velo} shows, a TB motion is always present at shear rates lower than approximately $\SI{0.1}{\second^{-1}}$, which is a new feature with respect to the deterministic phase diagram derived in Ref. \cite{lebe08}. 

      \begin{figure}
	\centering
	\includegraphics[width=0.99\linewidth]{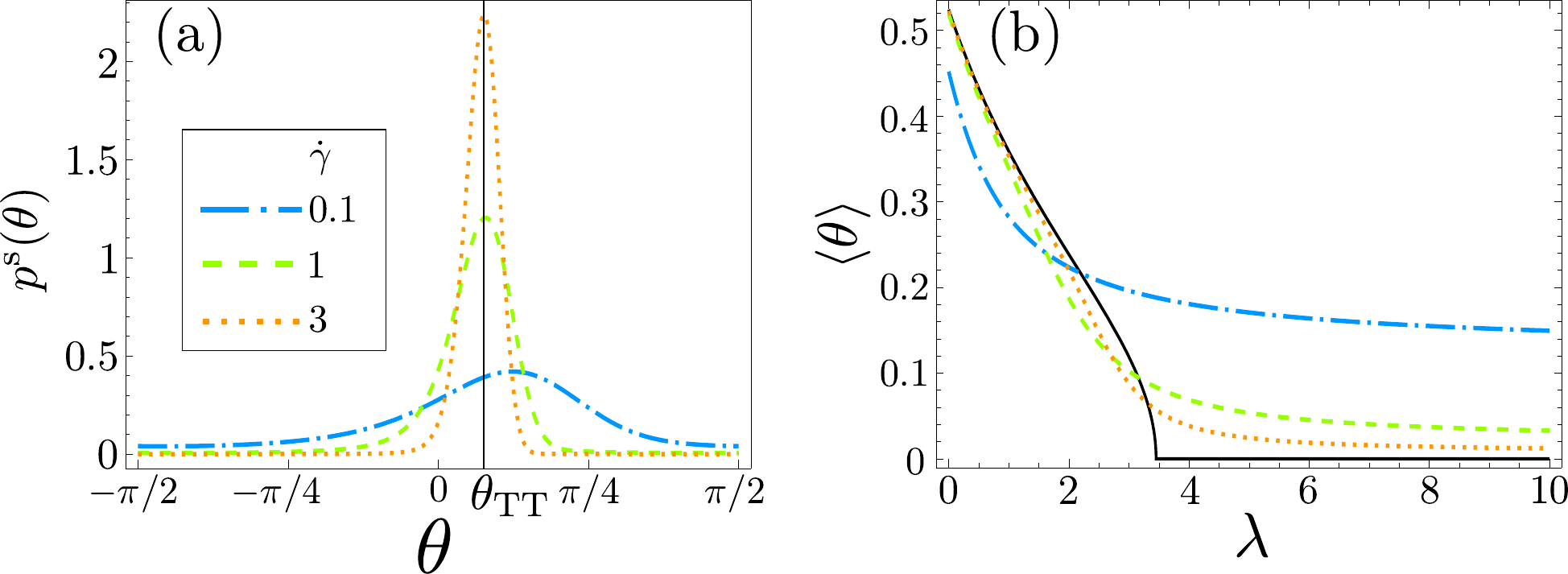} 
	\caption{(Color online) (a) Probability distribution of the orientation angle $\theta$ at $\lambda=2$ and different shear rates. (b) Mean inclination angle $\langle\theta\rangle$ as a function of the viscosity contrast. The solid black lines represent the KS model.}
	\label{ves_ps_lc}
      \end{figure}
    In order to illustrate this shear-induced phase transition, we plot the distribution function $p^\text{s}(\theta)$ at $\lambda=2$ and for $\dot\gamma=0.1,~1,~\text{and}~\SI{3}{\second^{-1}}$ in Fig. \ref{ves_ps_lc}(a). In the deterministic model, the vesicle tank treads at fixed orientation $\theta_\text{TT}$ (vertical line).  With thermal noise, the dynamics is different for these three shear rates indicated by crosses in Fig. \ref{ves_pd_velo}. For $\dot\gamma=\SI{0.1}{\second^{-1}}$, the vesicle's orientation cannot remain fixed because most fluctuations induce rotations and the regime is effectively TB ($\omega_\text{tu}=0.68$). The distribution is then relatively flat (dashed-dotted line). For $\dot\gamma=\SI{1}{\second^{-1}}$, the vesicle is tank treading and from time to time a full rotation is induced by fluctuations ($\omega_\text{tu}=0.13$). The probability distribution is almost symmetrically centered around $\theta_\text{TT}$ (dashed line). For $\dot\gamma=\SI{3}{\second^{-1}}$, the regime is effectively TT since tumbling practically never happens ($\omega_\text{tu}=0.0052$) and the orientation angle fluctuates around $\theta_\text{TT}$ due to thermal noise (dotted line). The mean inclination angle $\langle\theta\rangle$ as a function of $\lambda$ is also influenced by thermal noise - a fact already seen in Ref. \cite{nogu05} - and an exponential tail appears instead of sharp transition, as illustrated by Fig. \ref{ves_ps_lc}(b). Such a tail is similar to the one observed in a simulation including thermal noise \cite{me09}. A similar slow decay of the mean inclination angle has also been recently observed in an experiment \cite{kant06,zabu11}, but it seems that it can also be reproduced in deterministic simulations \cite{yazd12}.

\begin{figure}
	\centering
	\includegraphics[width=0.99\linewidth]{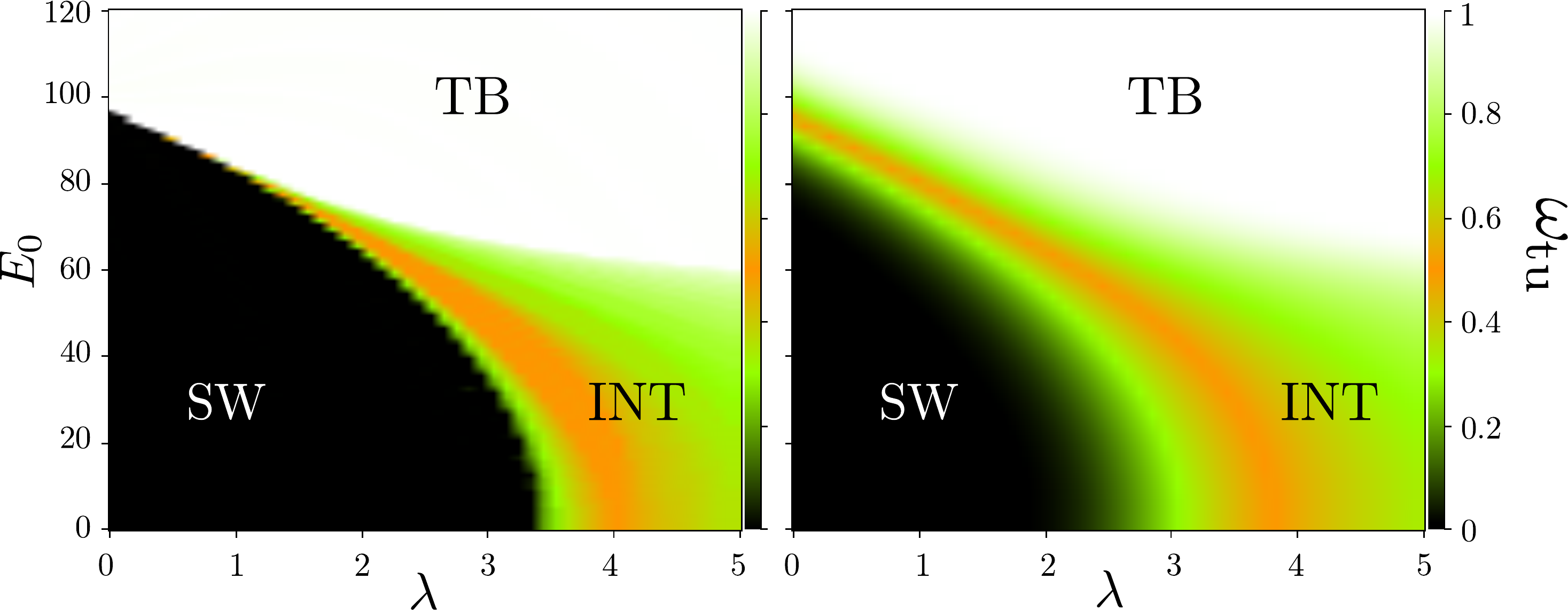}
	\caption{(Color online) Phase diagrams of the dynamics of capsules as a function of $\lambda$ and $E_0$ (in units of $k_\text{B}T$) for $\dot{\gamma}=\SI{1}{\second^{-1}}$ for the deterministic (left) and stochastic (right) case.}
	\label{caps_pd}
      \end{figure}

  {\sl Elastic capsules.--}
    For $E_0>0$, the solution of \eqref{fokkerplanck} cannot be separated and one has to solve the equation numerically. The phase diagrams of Fig. \ref{caps_pd} represent the tumbling ratio $\omega_\text{tu}$ as a function of the viscosity contrast $\lambda$ and the shape-memory energy $E_0$ at a shear rate of $\dot\gamma=\SI{1}{\second^{-1}}$. As for vesicles, $\omega_\text{tu}$ is never exactly $0$ or $1$ with thermal noise, but the values are extremely close in the black and white regions such that we can still effectively identify the SW and TB regimes, respectively. The region for which intermittent dynamics takes place is much wider in the presence of thermal noise. In particular, there is now a large intermittent region for  $\lambda\leq1$ which was not there in the deterministic case.

      \begin{figure}
	\centering
	\includegraphics[width=0.99\linewidth]{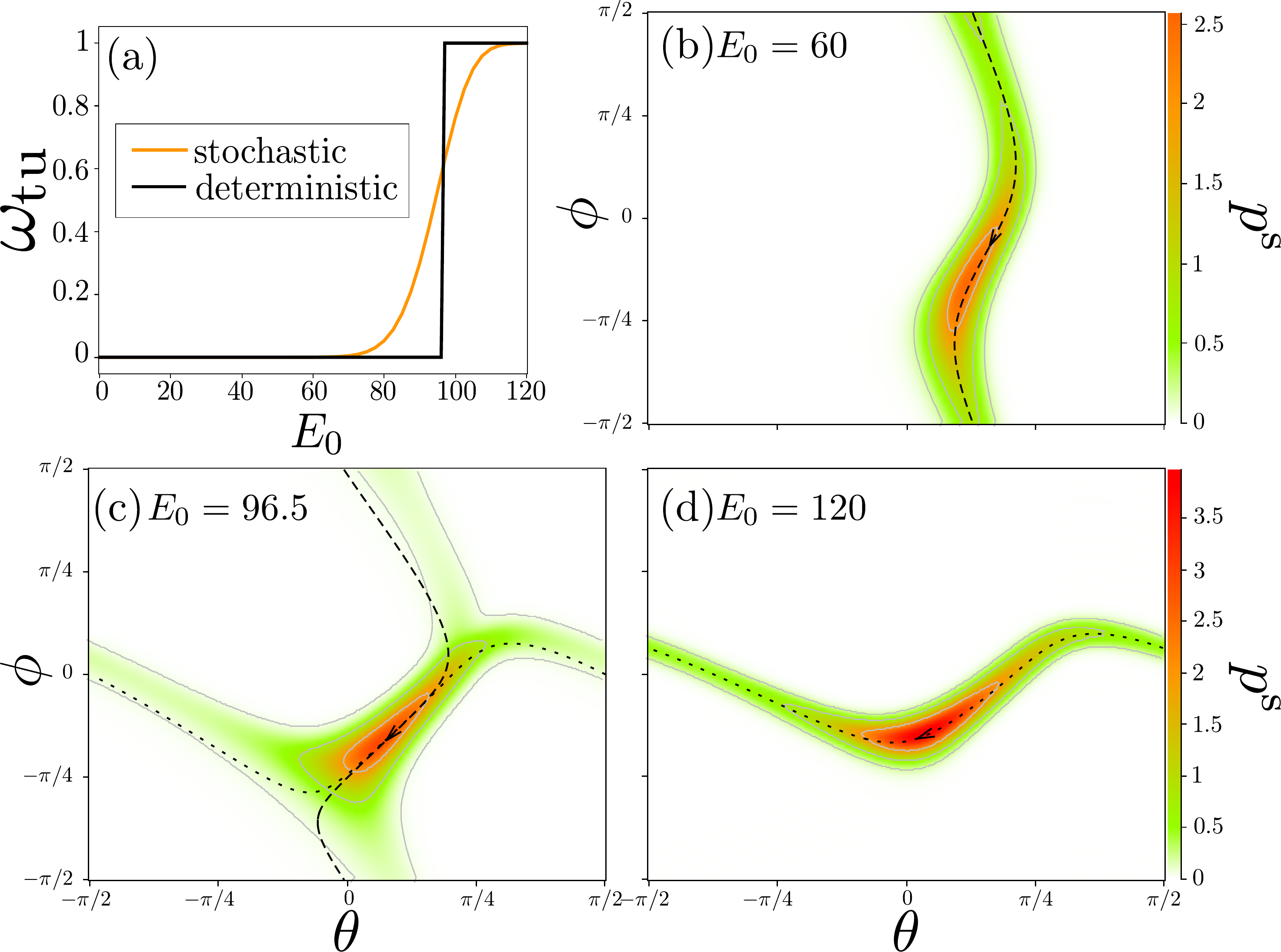}
	\caption{(Color online) (a) $\omega_\text{tu}$ as a function of $E_0$. (b)-(d) Stationary probability distribution $p^\text{s}(\theta,\phi)$ for three different values of $E_0$. The other parameters are $\lambda=0$ and $\dot\gamma=\SI{1}{\second^{-1}}$.}
	\label{caps_spec}
      \end{figure}
    This can be seen in Fig. \ref{caps_spec}, which shows the phase transition for $\lambda=0$. In the SS model, this transition is discontinuous and happens between $E_0=96$ (in units of $k_\text{B}T$), where the capsule is perfectly tank treading ($\omega_\text{tu}=1$), and $E_0=97$, at which perfect tumbling ($\omega_\text{tu}=0$) is observed [see Fig. \ref{caps_spec}(a)]. The thermal fluctuations turn the transition into a continuous one, which implies a mixing of both types of motion in a region of approximately $40 k_\text{B}T$ around the critical energy. This is illustrated by the stationary probability distribution at different values of the shape-memory energy. For $E_0=60$ [Fig. \ref{caps_spec}(b)], the capsule is in the TT regime and the probability distribution follows exactly the deterministic trajectory (dashed line), but the amplitude of the oscillations in $\theta$ is approximately two times larger. Similarly, for $E_0=120$ [Fig. \ref{caps_spec}(c)], the capsule is tank treading and the most probable path is the deterministic one (dotted line) with larger oscillations of $\phi$. At the transition [$E_0=96.5$, Fig. \ref{caps_spec}d], however, we observe a thermal mixing of the TT and TB regimes. The capsule exhibits both motions with a slight offset with respect to the deterministic paths at $E_0=96$ (dashed line) and $E_0=97$ (dotted line).

      \begin{figure}
	\centering
	\includegraphics[width=0.99\linewidth]{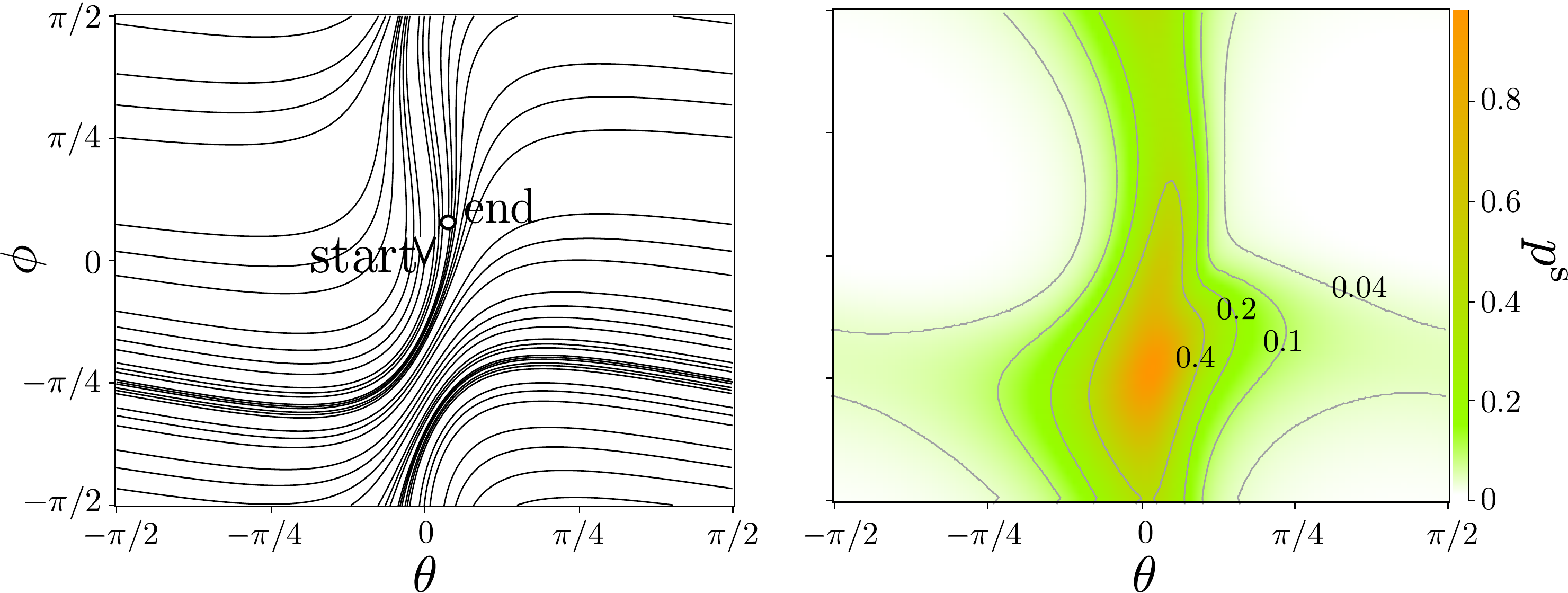}
	\caption{(Color online) Intermittent behavior with $\dot{\gamma}=\SI{1}{\second^{-1}}$, $\lambda=4$ and $E_0=40 k_\text{B}T$. Left: Deterministic trajectory of length $\SI{500}{\second}$. Right: Corresponding probability distribution in the presence of thermal noise.}
	\label{caps_4_40}
      \end{figure}
    In the presence of thermal noise, the intermittent regime predicted by the SS model thus still exists, but is superimposed with a thermal mixing of the dynamical regimes, as shown in Fig. \ref{caps_4_40}. The left panel shows one deterministic trajectory of the capsule in the $\theta-\phi$ phase space. The trajectory is quasi-periodic and consists of incomplete tumbling rotations (horizontal curves) interrupted by incomplete tank-treading motions since there must not be any intersection. The right panel shows the probability distribution in the presence of thermal noise. The tank-treading path is the most probable one, even though tumbling also happens. The white areas on the top right and top left are almost excluded, which was not the case for the deterministic trajectory. Complete TB rotations as well as complete TT rotations may occur since an intersecting trajectory is not forbidden for a stochastic motion.

  {\sl Concluding Perspectives.--}
    We have shown that under realistic conditions, the effects of thermal fluctuations on the dynamics of fluid vesicles and elastic capsules should be observable for objects having approximately the same dimensions as red blood cells in water. In most extant experiments on elastic capsules, either the capsules are too big \cite{kole12} or the outside medium has a too high viscosity \cite{abka07,dupi10} for showing thermal effects. Two important predictions concern the tumbling motion of fluid vesicles at low shear rates for all viscosity contrasts, and the thermal mixing of swinging and tumbling for elastic capsules at low viscosity contrast. In the future, the generalization of our model to deformable objects should be investigated, for which we expect that thermal noise would play an even more important role, especially for fluid vesicles as suggested by the experiments reported in Ref. \cite{zabu11}.

    We thank B. Lander for valuable help and acknowledge support from GIF.

\bibliography{/home/abreu/Documents/Literatur/doktor.bib}

\end{document}